\documentclass[twocolumn, pra, superscriptaddress]{revtex4-1}
\usepackage{amssymb}
\usepackage{amsmath}
\usepackage{epsfig}
\usepackage{graphics, graphicx}
\usepackage{bbold}
\usepackage{psfrag}
\usepackage{mathcomp}
\usepackage{subfigure}
\usepackage{verbatim}
\usepackage{color}
\usepackage[colorlinks,citecolor=blue]{hyperref}
\def\cp#1{\mathbf{#1}}

\begin{document}

\title{Enhancing the Efimov correlation in Bose polarons with large mass imbalance}
\author{Mingyuan Sun}
\affiliation{Institute for Advanced Study, Tsinghua University, Beijing, 100084, China}
\author{Xiaoling Cui}
\email{xlcui@iphy.ac.cn}
\affiliation{Beijing National Laboratory for Condensed Matter Physics, Institute of Physics, Chinese Academy of Sciences, Beijing, 100190, China}
\date{\today}

\begin{abstract}
We study the effect of Efimov physics (in the few-body sector) to the spectral response of Bose polaron, a many-body system consisting of an impurity immersed in a bath of bosonic atoms. 
We find that the Efimov correlation can be greatly enhanced by increasing the mass ratio between the bosons and the impurity, which results in visible signatures in the rf spectrum of the polaron. Using a diagrammatic approach up to the third-order virial expansion, we show  how the mass imbalance and the enhanced three-body effect modify the line shape and the width of the polaron spectrum. Moreover, we study the effect of a finite boson-boson interaction to the  spectrum. Taking the realistic system of Li impurities immersed in Cs bosons with a positive Cs-Cs scattering length,
we find a visible Efimov branch, which is associated with the second lowest Efimov trimer, in the polaron spectrum. In particular, by adjusting the boson density the Efimov branch can greatly hybridize with the attractive polaron branch leading to the spectrum broadening near their avoided level crossing. Our results can be directly probed in the cold atoms experiments on Li-Cs and Li-Rb Bose polarons. 
\end{abstract}
\maketitle

\section{Introduction}

Novel few-body correlations in the interacting many-body systems have intrigued great research interests in both condensed matter physics and also the field of ultracold atoms. The Efimov effect, characterized by an infinite number of trimer states near a two-body resonance and following the universal scaling law\cite{Efimov,Braaten}, represents one of the most intriguing three-body correlations in the quantum world. Given the successful explorations of Efimov physics in ultracold atoms\cite{Efimov_Exp0,Efimov_Exp1,Efimov_Exp1bu,Efimov_Exp3,Efimov_Exp4,Efimov_Exp5,Efimov_Exp9,Efimov_Exp10,Efimov_Exp6,Efimov_Exp7,Efimov_Exp8,Efimov_Exp11,rf_1,rf_2,scaling_1,scaling_2,scaling_3}, it is time to ask how such novel few-body correlation affects the property of a many-body system. To address this question, a simple yet non-trivial platform is the polaron system in ultracold gases, which consists of an impurity atom embedded in and interacting with a bath of fermionic or bosonic atoms, respectively called the Fermi or Bose polaron. Experimentally, the Fermi polaron\cite{Zwierlein,Salomon,Grimm,Kohl,Grimm2016,Roati} and the Bose polaron\cite{Aarhus,JILA,Lamb} have been successfully explored in ultracold gases, both of which exhibit the attractive and repulsive branches signifying the two-body correlations. Nevertheless, the signature of Efimov physics has not been reported in existing experiments, despite a number of theoretical proposals of dominating three-body correlations in polaron systems\cite{Parish, Zhou, Zinner1, Nishida, Cui1, Cui2, Levinsen1, Levinsen2, Giorgini}.

In a previous study, we pointed out that Efimov signatures can be visualized in Bose polarons with large mass imbalance\cite{Sun}. Such enhanced Efimov correlation has taken advantage of the following facts. First, compared to Fermi systems, the Bose system naturally more favors the formation of Efimov trimers due to the absence of Pauli principle\cite{Efimov, Braaten}. Secondly, for the hetero-nuclear atomic system with two bosons and a third distinguishable particle, the Efimov scenario can be greatly modified by their mass ratios\cite{Greene}. Specifically, a large mass ratio between the bosons and the third particle can give rise to deep ground state trimer and dense Efimov spectrum with small scaling factor. Thirdly, in the Bose polaron system, by tuning the boson-impurity interaction, the sizes of the Efimov trimers can become comparable to the inter-particle distance of the polaron system. In this case the Efimov trimers will have the optimized interference with the many-body background, thereby producing visible Efimov signatures in the polaron spectrum\cite{Sun}. 

In this paper, we will extend our previous work\cite{Sun} to investigate in detail how the three-body Efimov correlations affect the line shape and the width of the Bose polaron spectrum with different mass ratios. Using the diagrammatic approach up to the third order virial expansion, we can easily separate the three-body contributions from two-body ones and see clearly the isolated three-body effects in the spectrum of Bose polarons. It is found that as the mass ratio between the bosons and the impurity increases, the three-body Efimov correlations play more and more essential roles in the polaron spectrum. In addition, we study the effect of background boson-boson interaction to the polaron spectrum. Taking the realistic system of Li impurities immersed in Cs bosons with a positive Cs-Cs scattering length\cite{new_expt1, new_expt2} for example, 
we find a visible Efimov branch, which is associated with the second lowest Efimov trimer in the three-body sector, in the polaron spectrum. By adjusting the boson density, such Efimov branch can undergo an avoided level crossing with the attractive polaron branch, leading to much broadened spectra due to their hybridization near the crossing. 
These results demonstrate the enhanced Efimov correlations by mass imbalance in the many-body setting of Bose polarons, which hopefully can be probed in the current cold atoms experiments of Li-Cs and Li-Rb polaron systems. 

The rest of the paper is organized as follows. Section II is contributed to the general formalism of diagrammatic approach in studying the Bose polaron spectrum. In section III we apply the approach to the simplest case with no boson-boson interaction and discuss the effect of mass imbalance. In section IV we study the effect of finite boson-boson interaction, taking the realistic example of Li impurities immersed in Cs bosons with a positive Cs-Cs scattering length. Finally we discuss and summarize our results in section V. 

\section{General formalism}

We start from the Hamiltonian of Bose polaron system:
\begin{equation}
\mathcal{H}=\frac{{\bf p}_{i}^2}{2m_\text{i}}+\sum\limits_{j=1}^{N}\frac{{\bf p}_{b,j}^2}{2m_\text{b}}+U_{ib}\sum\limits_{j=1}^{N}\delta({\bf r}_i-{\bf r}_{b,j})+U_{bb}\sum\limits_{j<k}^{N}\delta({\bf r}_{b,j}-{\bf r}_{b,k}),
\end{equation}
where ${\bf r}_i$ and ${\bf p}_i$ are respectively the position and momentum of the impurity; ${\bf r}_{b,j}$ and ${\bf p}_{b,j}$ ($j=1,\dots,N$) are the position and momentum of $N$ identical bosons; $m_b$ and $m_i$ are respectively the masses of bosons and the impurity, and the mass ratio is denoted by $\eta\equiv m_b/m_i$; $U_{ib}$ ($U_{bb}$) is the bare coupling strength between bosons and the impurity (within bosons), which can be related to the s-wave scattering lengths $a_\text{ib}$ ($a_{bb}$). In this paper we consider $a_\text{ib}$ is highly tunable across resonance while $a_{bb}$ stays constant (zero or a finite value). 

In most of the existing cold atoms experiments\cite{Grimm,Kohl,Grimm2016,Roati,Aarhus,JILA}, an inverse radio-frequency(rf) spectroscopy has been carried out in obtaining the polaron spectrum, in which the resulted signal is directly proportional to the impurity spectral function at zero momentum\cite{Bruun}. In the following, we will derive this quantity from the diagrammatic approach in the framework of high-temperature virial expansion\cite{Kaplan, Leyronas1,Leyronas2, Hofmann1,Ngampruetikorn,Hofmann2}. The advantage of this approach is that it can incorporate the two-body and three-body contributions in a systematic and separable way, which allows us to directly extract the Efimov effect.

We start by expanding the free boson propagator in powers of the fugacity of bosons $z_b=e^{\beta\mu_b}$ ($\mu_\text{b}$ is the boson chemical potential and $\beta=1/(k_\text{B}T)$): 
\begin{eqnarray}
G^{(0)}(\mathbf{p},\tau)&=&e^{-(\epsilon_\mathbf{p}-\mu)\tau} [-\Theta(\tau)-n_b(\epsilon_\mathbf{p}-\mu_b)]\nonumber\\
&=&e^{\mu\tau}\sum_{n\geq 0} G^{(0,n)}(\mathbf{p},\tau)z_b^n, \label{G}
\end{eqnarray}
where $G^{(0,n)}(\mathbf{p},\tau)$ is $-\Theta(\tau) e^{-\tau\epsilon_\mathbf{p}}$ for $n=0$ and $-e^{-n\beta\epsilon_\mathbf{p}}e^{-\epsilon_\mathbf{p}\tau}$ for $n\geqslant1$; $\tau\in(0,\beta]$ is the imaginary time; $\epsilon_\mathbf{p}=p^2/(2m_b)$; $n_b(x)=1/(e^{\beta x}-1)$ is the Bose distribution function. 
Based on Eq.\ref{G}, all physical quantities of Bose polarons can be expanded in powers of $z_b$. Note that the fugacity of the impurity is sent to zero due to its negligible density (corresponding to the impurity chemical potential $\mu_i\rightarrow -\infty$). It is straightforward to check that the $n$-th order virial expansion will give rise to the impurity self-energy up to the order of $z_b^{n-1}$. 

\begin{figure}[t] 
\includegraphics[width=9cm]{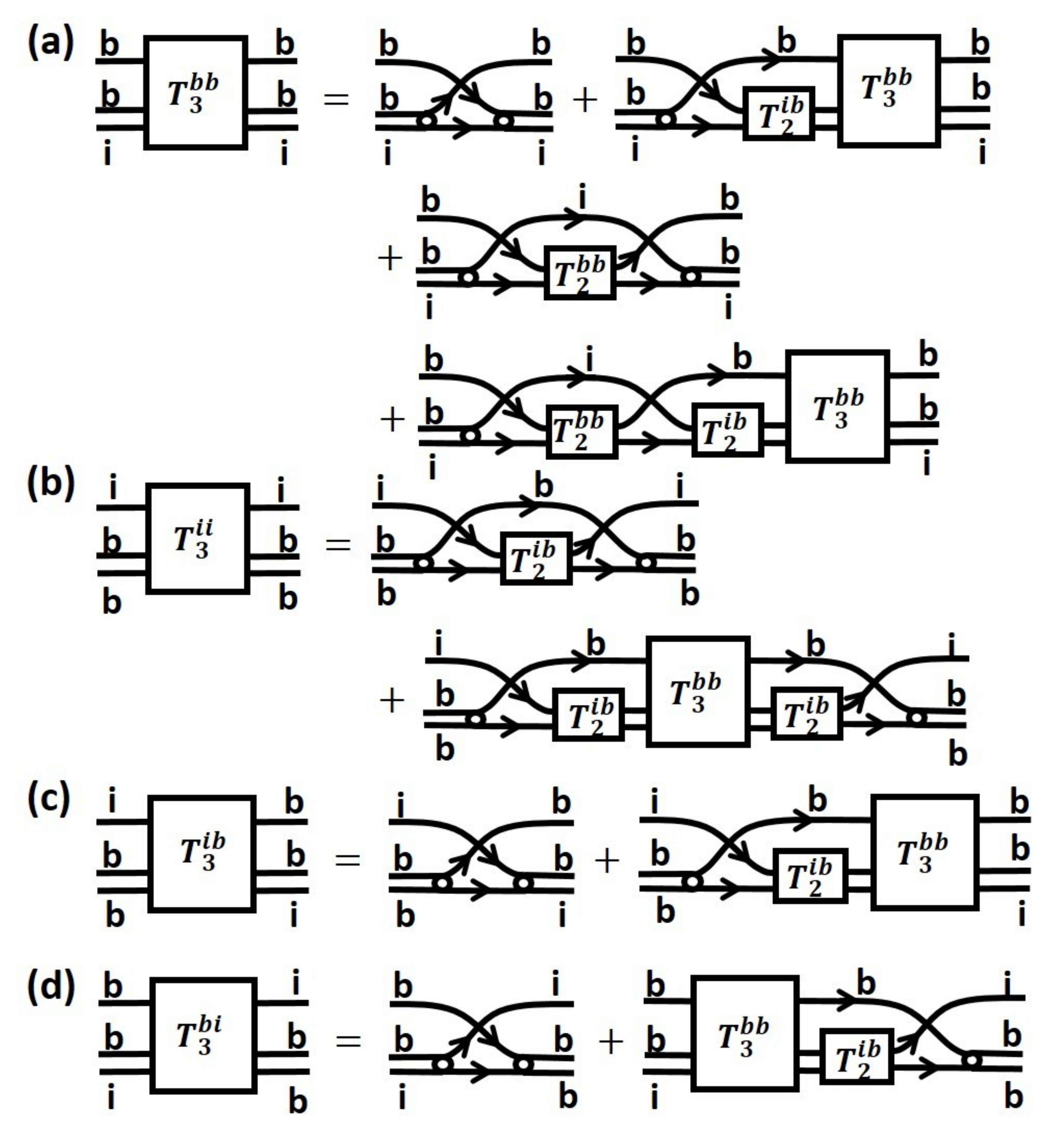}
\caption{Feynman diagrams for the atom-dimer scattering matrix $T_3^{\alpha\beta}$. Here $\alpha,\beta=i$ or $b$ is the shorthand for impurity or boson. (a,b,c,d) respectively correspond to $\alpha\beta=bb, ii, ib, bi$. }\label{fig1}
\end{figure}

Before going to the diagrams of the impurity self-energy, let us first introduce the two-body scattering matrix $T_2^{\alpha\beta}$ and the atom-dimer scattering matrix $T_3^{\alpha\beta}$. Namely, $T_2^{\alpha\beta}$ denotes the two-body scattering matrix between $\alpha$ and $\beta$, with $\alpha\beta=ib$ or $bb$ for the shorts of impurity-boson or boson-boson pair:    
\begin{equation}
T_2^{\alpha\beta}=\frac{2\pi}{m_{\alpha\beta}} \frac{1}{a_{\alpha\beta}^{-1}-\sqrt{-2m_{\alpha\beta} E}}, 
\end{equation}  
where $E$ is the scattering energy, and $m_{ib}=\frac{m_bm_i}{m_b+m_i}$, $m_{bb}=m_b/2$ are the reduced masses. 
$T_3^{\alpha\beta}$ is the atom-dimer scattering matrix with the incoming atom denoted by $\alpha$ and outgoing atom denoted by $\beta$, and there are four combinations $\alpha\beta=bb,ii,ib,bi$ as shown in Fig.\ref{fig1}(a-d). Among all of them, $T_3^{bb}$ can be first obtained from the diagrams in Fig.\ref{fig1}(a), which reads: 
\begin{widetext}
\begin{eqnarray}
  T_{3}^{bb}(\mathbf{p_1},\mathbf{p_2},E)&=&\frac{1}{E-\epsilon_{\cp p_1}-\epsilon_{\cp p_2}-(\cp p_1+\cp p_2)^2/(2m_i)}+\int\frac{d^3 \mathbf{q}}{(2\pi)^3}\frac{T_{2}^{ib}(E-\frac{q^2}{2m_{AD}})}{E-\epsilon_{\cp p_1}-\epsilon_{\cp q}-\frac{(\cp p_1+\cp q)^2}{2m_i}} T_{3}^{bb}(\mathbf{q},\mathbf{p_2},E) \nonumber \\
    &&+\int\frac{d^3 \mathbf{q}}{(2\pi)^3}\frac{T_{2}^{bb}(E-\frac{q^2}{2m_{iD}})}{E-\epsilon_{\cp p_1}-\epsilon_{\cp p_1+\cp q}-\frac{{\cp q}^2}{2m_i}} \frac{1}{E-\epsilon_{\cp q+\cp p_2}-\epsilon_{\cp p_2}-\frac{{\cp q}^2}{2m_i}} \nonumber \\
    &&+\int\frac{d^3 \mathbf{q}}{(2\pi)^3} \int\frac{d^3 \mathbf{q_1}}{(2\pi)^3} \frac{T_{2}^{bb}(E-\frac{q^2}{2m_{iD}})}{E-\epsilon_{\cp p_1}-\epsilon_{\cp p_1+\cp q}-\frac{{\cp q}^2}{2m_i}} \frac{T_{2}^{ib}(E-
    \frac{q_1^2}{2m_{AD}})}{E-\epsilon_{\cp q+\cp q_1}-\epsilon_{\cp q_1}-\frac{{\cp q}^2}{2m_i}} T_{3}^{bb}(\mathbf{q_1},\mathbf{p_2},E)
\end{eqnarray}
\end{widetext}
where $E$ is the scattering energy, and $\mathbf{p_1},\mathbf{p_2}$ are respectively the relative momenta of the incoming and outgoing atom-dimer states in the center-of-mass frame. $m_{AD}=m_b(m_b+m_i)/(2m_b+m_i)$ and $m_{iD}=2m_i m_b/(2m_b+m_i)$ are respectively the boson's and the impurity's reduced mass for the atom-dimer scattering. Then the remaining matrixes can be straightforwardly expressed by $T_3^{bb}$ and $T_2^{ib},\ T_2^{bb}$ as shown in Fig.\ref{fig1}(b-d). 

\begin{widetext}

\begin{figure}[h] 
\includegraphics[width=15cm]{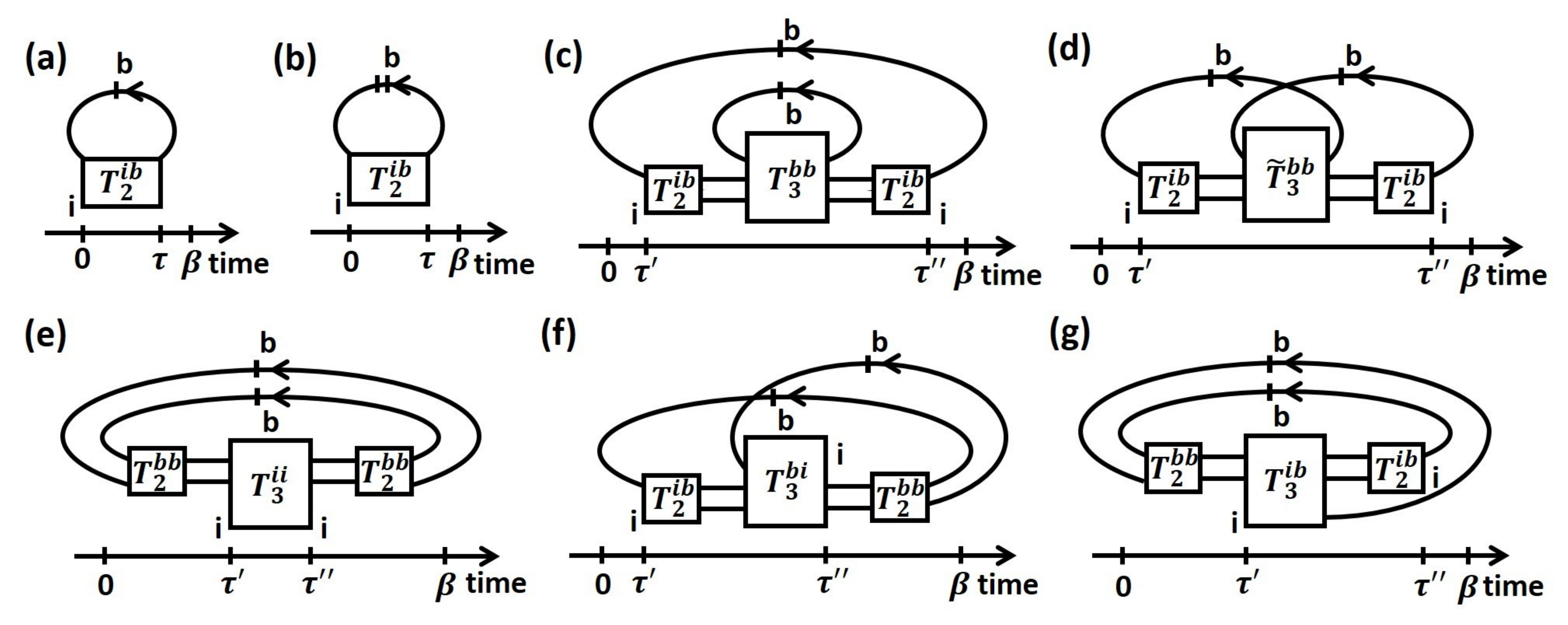}
\caption{Feynman diagrams of the impurity self energy $\Sigma^{(1)}$ (a) and $\Sigma^{(2)}$(b-g). $\tilde{T}_3$ is $T_3$ excluding the first Born term. The boson propagator line with $n$ vertical dashes denotes the $n$-th order contribution $G^{(0,n)}$ in Eq.\ref{G}.}\label{fig2}
\end{figure}

\end{widetext}

In Fig.\ref{fig2} we plot the Feynman diagrams for the impurity self energy $\Sigma({\bf k},\tau)$ up to the order of $z_b^2$, i.e., in the framework of the third-order virial expansion where the two-body and three-body contributions are fully taken into account. Compared to Ref.\cite{Sun}, here additional diagrams are included in Fig.\ref{fig2} (e,f,g) due to the presence of a finite boson-boson interaction.

Fig.\ref{fig2}(a) leads to the lowest order of $\Sigma$ in $z_b$:  
\begin{equation}
\Sigma^{(1)}=z_b \int\frac{d^3\mathbf{P}}{(2\pi)^3} e^{-\beta \epsilon_{\mathbf{P-k}}} T_2^{ib}\left(\omega+i\delta+\epsilon_{\mathbf{P-k}}-\frac{P^2}{2(m_i+m_b)}\right);
\end{equation}
The remaining diagrams (b-g) give rise to the second order contributions $\Sigma^{(2)}\sim z_b^2$. The sum of the diagrams in (b,c,d)  gives:
\begin{widetext}
\begin{eqnarray}
 \Sigma^{(2; bcd)}&=&z_b^2 \left\{\int\frac{d^3\mathbf{P}}{(2\pi)^3}  e^{-2\beta \epsilon_{\mathbf{P-k}}}T_2^{ib}\Big(\omega+i\delta+\epsilon_{\mathbf{P-k}}-\frac{P^2}{2(m_i+m_b)}\Big)  \right. \nonumber\\
   &&+\int\frac{d^3\mathbf{p_1}}{(2\pi)^3}\int\frac{d^3\mathbf{p_2}}{(2\pi)^3} e^{-\beta(\epsilon_{\mathbf{p_1}}+ \epsilon_{\mathbf{p_2}})} \left[T_2^{ib}\Big(\omega+i\delta+\Delta-\frac{{p_1'}^2}{2m_{AD}}\Big)\right]^2 T_3^{bb}(\mathbf{p'_1},\mathbf{p'_1},\omega+i\delta+\Delta) \nonumber\\
   &&+\left. \int\frac{d^3\mathbf{p_1}}{(2\pi)^3}\int\frac{d^3\mathbf{p_2}}{(2\pi)^3}  e^{-\beta(\epsilon_{\mathbf{p_1}}+ \epsilon_{\mathbf{p_2}})}  T_2^{ib}\Big(\omega+i\delta+\Delta-\frac{{p'_1}^2}{2m_{AD}}\Big)\tilde{T}_3^{bb}(\mathbf{p'_1},\mathbf{p'_2},\omega+i\delta+\Delta) T_2^{ib}\Big(\omega+i\delta+\Delta-\frac{{p'_2}^2}{2m_{AD}}\Big) \right\}.
\end{eqnarray}
\end{widetext}
Here $\Delta=\epsilon_{\mathbf{p_1}}+\epsilon_{\mathbf{p_2}}-P_t^2/(2M)$, with $\mathbf{P_t}=\mathbf{k}+\mathbf{p_1}+\mathbf{p_2}$ and $M=2m_b+m_i$ respectively the total momentum and the total mass of three-body system; $\mathbf{p'_{1,2}}=\mathbf{p_{1,2}}-m_b\mathbf{P_t}/M$ and $m_{AD}=m_b(m_b+m_i)/M$ are the relative momenta and the reduced mass for atom-dimer scattering. 

The sum of the diagrams in (e,f,g) give:
\begin{widetext}
\begin{eqnarray}
  \Sigma^{(2,efg)}&= & z_b^2 \int\frac{d^3\mathbf{P}}{(2\pi)^3} \int\frac{d^3\mathbf{p_1}}{(2\pi)^3}\int\frac{d^3\mathbf{p_2}}{(2\pi)^3}  \int_0^\infty dx e^{-\beta (x+\frac{P^2}{4m_b})} \rho (x)  \left[ T_{2}^{ib}\Big(\omega+i\delta+\Delta-\frac{{p_1'}^2}{2m_{AD}}\Big) \cdot (2\pi)^3 \delta(\mathbf{p_1-p_2}) \right. \nonumber\\
 && \left. 
   +T_{2}^{ib}\Big(\omega+i\delta+\Delta-\frac{{p'_1}^2}{2m_{AD}}\Big) T_{3}^{bb}(\mathbf{p'_1},\mathbf{p'_2},\omega+i\delta+\Delta) T_{2}^{ib}\Big(\omega+i\delta+\Delta-\frac{{p'_2}^2}{2m_{AD}}\Big) \right] \label{efg}
\end{eqnarray}
with 
\begin{equation}
  \rho (x)= -\frac{1}{\pi}Im \left[ \frac{T_{2}^{bb}(x+i0^+)}{(x+i0^+-\varepsilon_1)(x+i0^+-\varepsilon_2)} \right]
\end{equation}
\end{widetext}
Here $\varepsilon_1=\epsilon_{\mathbf{p_1}}+\epsilon_{\mathbf{P-p_1}}-P^2/(4m_b)$; $\varepsilon_2=\epsilon_{\mathbf{p_2}}+\epsilon_{\mathbf{P-p_2}}-P^2/(4m_b)$; $\Delta=x+P^2/(4m_b)-P_t^2/(2M)$, and $\mathbf{P_t}=\mathbf{k}+\mathbf{P}$ is the total momentum of three-body system. In the case of a positive $a_{bb}$,  we have neglected the contribution from the bound state of two bosons. 

By summing up all the diagrams in Fig.\ref{fig2}, we obtain the impurity self-energy $\Sigma=\Sigma^{(1)}+\Sigma^{(2)}$ up to the order of $z_b^2$, which has included all the two-body and three-body contributions. The spectral function can be computed from the propagator of the impurity, $G_i({\cp k},\omega)=(\omega+i\delta-k^2/(2m_i)-\Sigma({\cp k},\omega+i\delta))^{-1}$, as
\begin{equation}
A({\cp k},\omega)=-\frac{1}{\pi} {\rm Im}\Big(G_i({\cp k},\omega)\Big). \label{A}
\end{equation}

\section{Efimov correlation enhanced by mass imbalance}

In this section, we focus on the effect of mass imbalance to the spectral response of Bose polarons. To facilitate the discussion, here we consider the simple case when the background boson-boson interaction is absent. Therefore among the diagrams of $T_3$ in Fig.\ref{fig1} only $T_3^{bb}$ is non-zero, and in Fig.\ref{fig2} only the diagrams of (a-d) are relevant. The specific effect of a finite boson-boson interaction will be discussed in Section IV, and we emphasize here that the general concept of using mass imbalance to enhance the Efimov correlation in Bose polarons will not be affected by the background boson-boson interaction.  

As pointed out in our earlier study\cite{Sun}, to facilitate the visibility of Efimov signatures in polaron system, one requires the size of the Efimov trimer($l_t$) be comparable to the inter-particle distance($d$) of the underlying many-body system. This is because if the trimers are too shallow ($l_t\gg d$), as in the recent Bose polaron experiments\cite{Aarhus, JILA}, the three-body correlation can be easily washed out by the two-body ones near the resonance. In the opposite limit, if the trimers are too deep ($l_t\ll d$), the Efimov signal is also weak in the polaron spectrum as measured from the inverse rf spectroscopy, because  such deep trimers have little wave function overlap with the initial scattering state. Thus the optimal situation is  $l_t\sim d$, and to achieve this condition one would need the mass imbalance to tune the Efimov scenario in hetero-nuclear atomic systems. 

\begin{widetext}

\begin{figure}[t] 
\includegraphics[height=11cm]{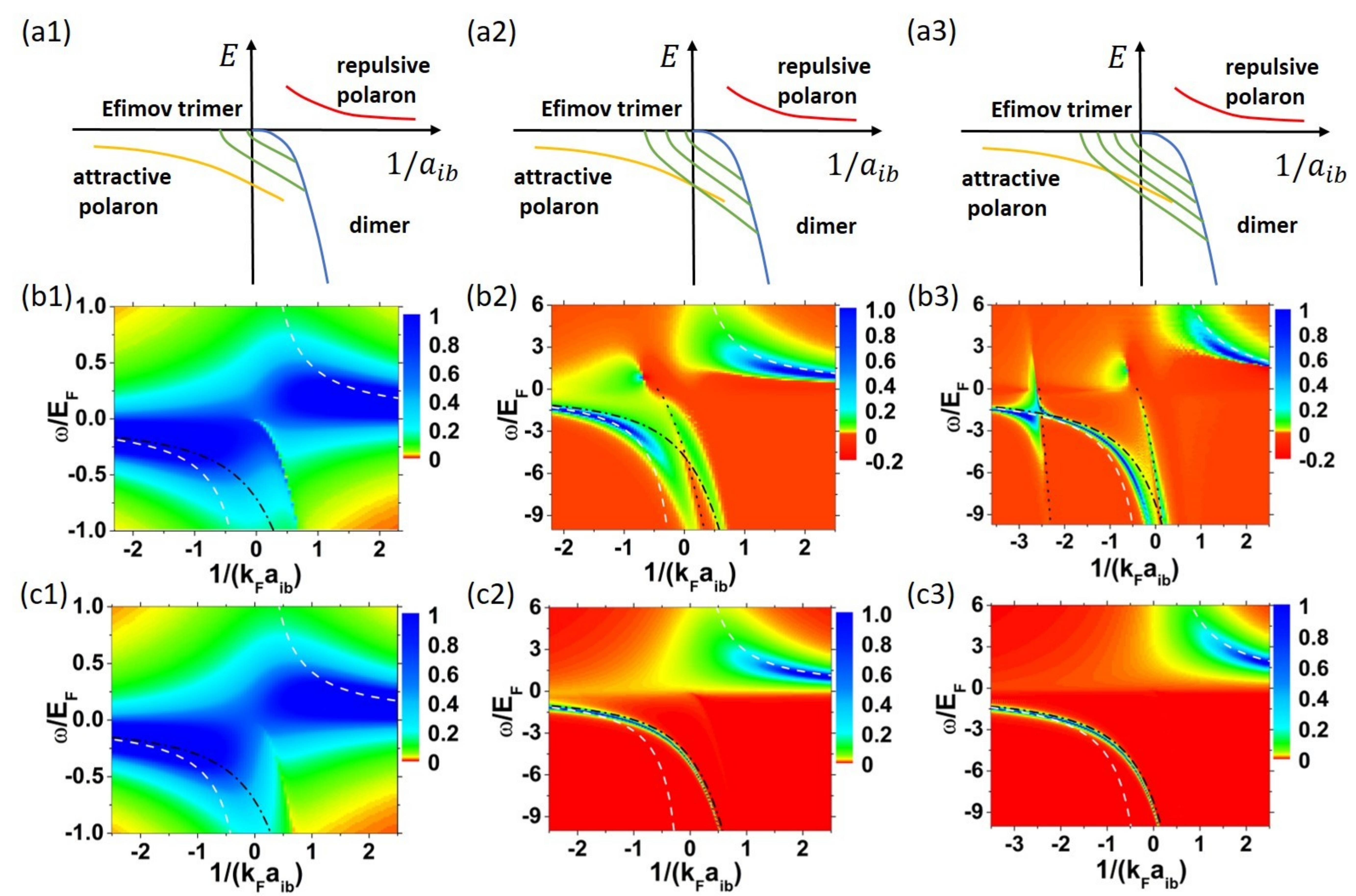}
\caption{(Color Online). Schematics of the Efimov states with respect to the polaron branches (first row), and the contour plots of impurity spectral function $A(0,\omega)$ by taking into account all the two-body and three-body contributions (second row) or by only considering the two-body contribution (third row). Three systems with different impurity-boson combinations are considered (from the first to the third column): (I)$^{39}$K-$^{39}$K, (II)$^{7}$Li-$^{87}$Rb and (III)$^{6}$Li-$^{133}$Cs. In our numerical calculations, we take the boson fugacity $z_b=0.1$, and the boson density $n_b=2\times 10^{14}$cm$^{-3}$ for all systems, which determines $k_F=(6\pi^2n_b)^{1/3}$ and $E_F=k_F^2/(2m_b)$. The scattering lengths for the first Efimov resonance in different systems are taken from Ref.\cite{Aarhus, Li-Rb, scaling_2,scaling_3}. In the spectrum, we also show the 
trimer energy (black dotted), and the polaron energy from mean-field theory (white dashed) and from the variational approach up to single boson excitations\cite{LiWeiran} (black dash-dotted). Here we assume no boson-boson interaction for all systems. The plots of (b1,b3) are the same as shown in Ref.\cite{Sun}. }\label{fig3}
\end{figure}

\end{widetext}

To illustrate the idea, we consider three experimentally well-studied systems with different impurity-boson combinations: (I) $^{39}$K-$^{39}$K\cite{Aarhus}, (II) $^{7}$Li-$^{87}$Rb\cite{Li-Rb} and (III) $^{6}$Li-$^{133}$Cs\cite{scaling_2,scaling_3}, which respectively give the mass ratio $\eta=1,\ 12.4$ and $22.2$. In Fig.\ref{fig3} (a1-a3), we show the schematics of Efimov energy levels with respect to the attractive/repulsive polaron branches in these systems.  The effect of mass imbalance, as discussed previously, is to modify the Efimov scenarios in the three-body system of two bosons and the third particle\cite{Braaten, Greene}. Namely, by increasing $\eta$ the Efimov scaling can be greatly reduced\cite{Braaten} and the ground state Efimov trimer appears far from resonance and can be rather deep at resonance\cite{Greene}. As shown in Fig.\ref{fig3} (a1-a3), with increasing $\eta$ from (a1) to (a3), the Efimov spectra become gradually denser and the energy levels move downward to get close or even level cross with the attractive polaron branch (see (a2) and (a3)). Thus one can expect that the Efimov signature will not show up in (a1) (system I) because of $l_t\gg d$ but could be visible in (a2,a3) (systems II,III) given $l_t\sim d$. 


The basic idea illustrated above can be verified by the numerical calculation of the impurity spectral function $A(k=0,\omega)$ based on the formalism presented in Section II. In Fig.\ref{fig3}(b1-b3), we show the spectrum of different polaron systems by taking into account all the two-body and three-body contributions (by summing up the diagrams in Fig.\ref{fig2}(a-d)). To separate out the three-body effect, in Fig.\ref{fig3}(c1-c3), we further show the results by only considering the two-body contributions (diagrams in Fig.\ref{fig2}(a,b)). Here we take the boson fugacity $z_b=0.1$ and a uniform boson density $n_b=2\times 10^{14}$cm$^{-3}$ for all different systems. We use $k_F^{-1}=(6\pi^2n_b)^{-1/3}$ and $E_F=k_F^2/(2m_b)$ as the units of length and energy. All systems are in the high-temperature regime with $T/E_F=3.75$. 

Clearly, we see that for system (I) with small mass ratio $\eta=1$, only the attractive and repulsive polaron branches are visible but not any signature of Efimov physics (Fig.\ref{fig3}(b1)). Meanwhile, by comparing (b1) and (c1) we can see that the two-body contributions dominate in the resulted spectrum while three-body ones take little effect. All these results are consistent with our expectation on the rather shallow Efimov trimers with size $l_t\gg d$. On the contrary, for systems (II) and (III) with large $\eta$, the three-body (Efimov) contributions play an important role in the resulted spectrum. For system (II),  Fig.\ref{fig3} (b2) shows an additional Efimov branch besides the ordinary attractive and repulsive polaron branches. Such Efimov branch is associated with the first Efimov trimer emerging at $1/(k_F a_-^{(1)})=-0.44$. As tuning $1/a_{ib}$ this branch can undergo an avoided level crossing with the attractive branch, giving rise to the spectral broadening due to the enhanced hybridization between these branches. For system (III) with larger $\eta$, because of the even deeper and denser Efimov states, we can see in Fig.\ref{fig3}(c2) two visible Efimov branches in the spectrum, which are respectively associated with the lowest two Efimov trimers emerging at $1/(k_Fa_-^{(1)})=-2.56$ and $1/(k_Fa_-^{(2)})=-0.40$. As tuning $1/a_{ib}$ these Efimov branches can either undergo an avoided level crossing or get close to the attractive polaron branch, which both lead to the spectral broadening near the regime of inter-branch hybridization. 

The three-body contributions in systems (II) and (III) are remarkable by comparing Fig.\ref{fig3} (b2,c2) with (b3,c3). First, the three-body effect directly leads to the appearance of Efimov branches. Namely, such branches can only show up in (b2,c2) which incorporate the three-body effect but not in (b3,c3). Secondly, with three-body effect, the spectral width of the attractive branch in (b2,c2) is reasonably broader than that in (b3,c3). This can be attributed to the deep Efimov trimers below the attractive branch, which provide an additional channel for the latter to decay. Finally, we see the three-body effect can produce additional signals in (b2,c2) at positive frequency $\omega>0$. This can be related to the effect of Efimov resonance, i.e., when the Efimov trimers start to emerge from the scattering threshold, to the scattering states of polarons. Nearby zero frequency, we also find a small unphysical parameter region with negative spectral function, and we attribute this to the convergence problem of virial expansion and the absence of higher-order contributions in the self-energy. 

Fig.\ref{fig3}(b1-b3) and (c1-c3) also reveal another notable effect of the mass imbalance, namely, by increasing $\eta$ the attractive and repulsive branches acquire much narrower relative spectral width, which is defined by the ratio of the absolute width to the mean location of the spectral peak. This has been pointed out earlier in Ref.\cite{Sun} where three-body effect was taken into account, while here we show that this statement equally applies if only consider the two-body contributions (see (c1-c3). Therefore this feature does not rely on the order of virial expansion, but rather an intrinsic property uniquely produced by the mass imbalance. As we have taken the same boson fugacity $z_b$ for all systems, which gives the same ratio between the thermal wavelength ($\lambda_T$) and inter-particle distance ($d$), this observation suggests that for a given $\lambda_T/d$, the Bose polaron quasi-particle is more well-defined for larger mass ratio $\eta$.

We note that previous studies\cite{Levinsen2, Giorgini} have revealed the Efimov effect to the energetics of the Bose polarons with relatively small mass imbalance. In comparison, we study the Efimov signatures in the spectral response of Bose polarons facilitated by the large mass imbalance. Thus the setting of our work is different from the previous studies\cite{Levinsen2, Giorgini}. There is, however, an intrinsic relation between our work and Ref.\cite{Levinsen2}, in that the avoided level crossing between the Efimov branch and the attractive polaron branch in this work (as shown in Figs.\ref{fig3},\ref{fig4}) is consistent with the physics of  atom-trimer continuity in the ground state of the Bose polarons as pointed out in Ref.\cite{Levinsen2}.

\section{Effect of a finite boson-boson interaction}

In this section we discuss the effect of a finite boson-boson interaction with $a_{bb}\neq 0$. In the presence of a finite $a_{bb}$, all the four $T_3^{\alpha\beta}$ in Fig.\ref{fig1} should be modified from zero $a_{bb}$ case and all the self-energy diagrams in Fig.\ref{fig2} involving $T_3$ and $T_2^{bb}$ should be affected. Specifically, compared to zero $a_{bb}$ case, Fig.\ref{fig2}(c,d) produce different results due to the change of $T_3^{bb}$, and Fig.\ref{fig2}(e,f,g) also give non-zero contributions. 

\begin{figure}[h] 
\includegraphics[width=9cm]{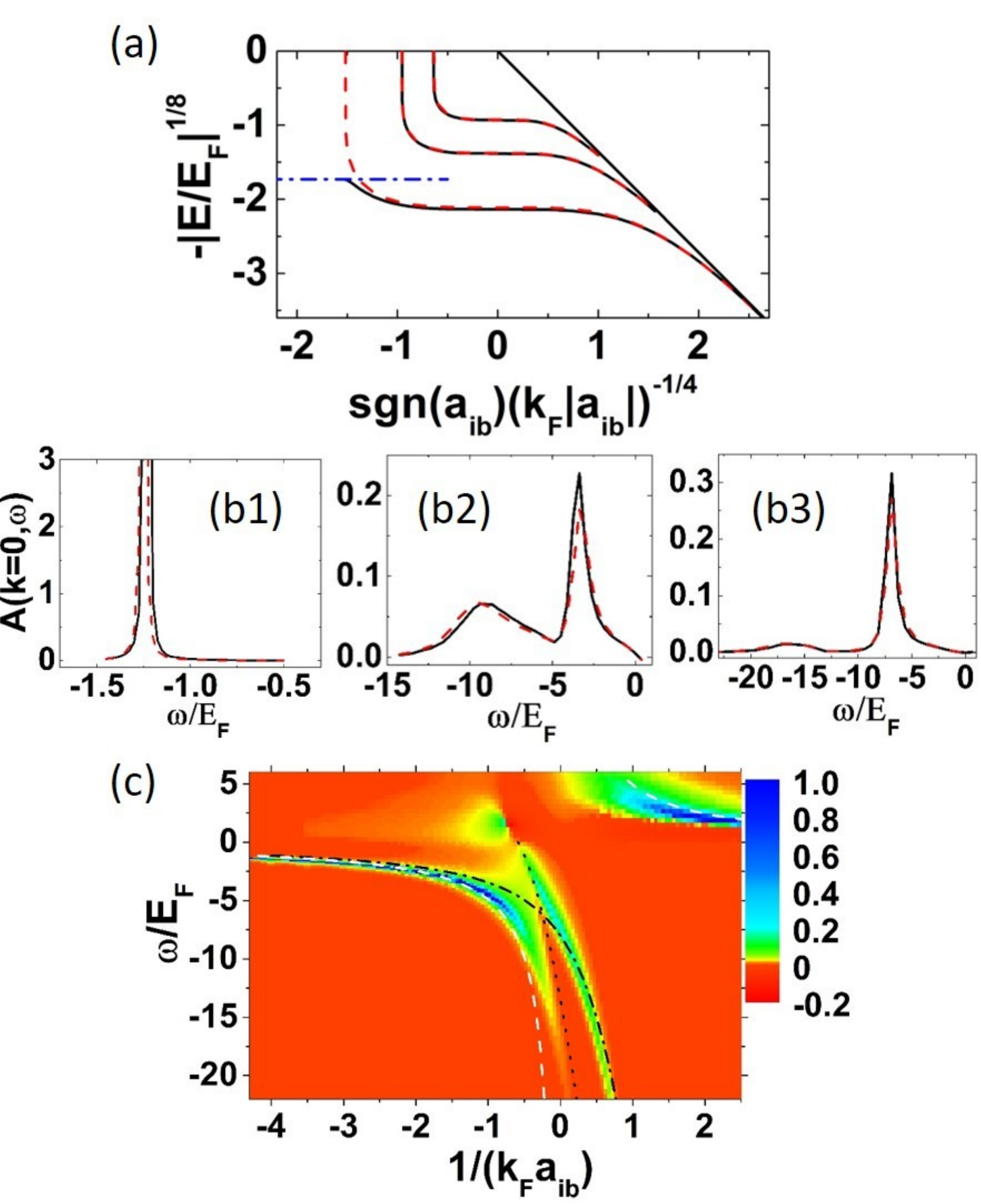}
\caption{(Color Online). Efimov spectrum in a three-body system (a), and the Bose polaron spectrum (b1-b3, c) for $^{6}$Li-$^{133}$Cs system with a finite boson-boson scattering length $a_{bb}=10$nm. We take $z_b=0.1$ and the boson density $n_b=0.64\times10^{14}$ cm$^{-3}$. In (a), we show the three lowest Efimov states of Li-Cs-Cs system with zero (red dashed) and finite $a_{bb}$ (black solid). The blue dash-dotted line shows the boson dimer energy supported by the positive $a_{bb}$. In (b1-b3) we show the slices of $A(0,\omega)$ at different interaction strength $1/(k_Fa_{ib})=-3.75,-0.3,0$. The black solid (red dashed) curves are obtained with (without) the contributions from (e,f,g) diagrams in Fig.\ref{fig2} (or Eq.\ref{efg}). In the contour plot (c), different lines are the same as in Fig.\ref{fig3}. }\label{fig4}
\end{figure}

Here we consider the $^{6}$Li-$^{133}$Cs system near $889$G Feshbach resonance, where the Cs-Cs scattering length is $a_{bb}=10$nm. 
Recent theoretical and experimental studies have shown that the presence of a positive $a_{bb}$ can greatly modify the Efimov scenario\cite{new_expt1, new_expt2}, in that the ground state trimer of Li-Cs-Cs system merges into the Cs-Cs dimers as tuning the Li-Cs scattering length which makes the first Efimov resonance disappear. 

In Fig.\ref{fig4}(a), we numerically verify above statement by calculating the lowest three trimer energies $E_t^{(n)}\ (n=1,2,3)$ from the pole of any $T_3$ matrix in Fig.\ref{fig1}. We find that indeed the first trimer state ($E_t^{(1)}$) merges into the Cs-Cs dimer line ($E_{2,bb}=-1/(m_ba_{bb}^2)$) instead of touching the scattering threshold. However, for other trimer states with index $n\ge 2$, the Efimov spectra are hardly modified by $a_{bb}$. This can be attributed to the well-separated length scales, i.e.,  $a_{bb}\ll |a_-^{(n)}|, l_t^{(n)}$ for $n\ge 2$, where $a_-^{(n)}, l_t^{(n)}$ are respectively the scattering length for the $n$-th Efimov resonance and the size of the $n$-th trimer. We thus expect that one can still utilize the second lowest Efimov trimer to visualize the Efimov signature in Li-Cs Bose polarons. 

In Fig.\ref{fig4} (b1-b3) we show slices of $A(0,\omega)$ for Li-Cs polaron system at three typical interaction strength $1/(k_Fa_{ib})=-3.75, -0.3$ and $0$. Here we again take $z_b=0.1$ but choose a smaller boson density $n_b=0.64\times10^{14}$ cm$^{-3}$ compared to that in Fig. \ref{fig3}. The purpose of a smaller $n_b$ here is to reduce the energy of the attractive branch so as to enable the maximized hybridization with the Efimov branch in the form of the (avoided) level crossing. Indeed, we find no Efimov signature near $a_-^{(1)}$ (Fig.\ref{fig4}(b1)); while near $a_-^{(2)}$, a sharp Efimov peak appears in the spectrum and it coexists with the attractive polaron peak (Fig.\ref{fig4}(b2));  further tune away from the avoided level crossing, the Efimov peak evolves to a single peak signifying the attractive polaron (Fig.\ref{fig4}(b3)). When plotting this figures, we show the results both with and without including the contribution from (e,f,g) diagrams in Fig.\ref{fig2}, and find that they do not make too much difference.  Therefore in this  case the (e,f,g) diagrams take little effect in the resulted spectrum, and for simplicity they have been neglected in producing the contour plot in Fig.\ref{fig4} (c). 

In Fig.\ref{fig4} (c) we show contour plot of the spectrum in the ($1/(k_Fa_{ib}),\ \omega$) parameter plane. As expected we see an avoided level crossing between the Efimov branch (associated with the second lowest Efimov trimer) and the attractive polaron branch. Similar to  Fig.\ref{fig3}, near the region of their avoided level crossing the spectra are greatly broadened due to the enhanced inter-branch hybridization with similar energies. Comparing Fig.\ref{fig4} (c) and Fig.\ref{fig3} (b3), we see that the relative energy between different branches can be conveniently tuned by the density of boson system.

In this section we have discussed the effect of a finite $a_{bb}$ to the spectrum of Li-Cs Bose polarons. When generalized to Li-Rb system\cite{Li-Rb}, since the Rb-Rb scattering length $a_{bb}\sim 5$nm is much smaller than $|a_-^{(1)}|\sim 100$nm, we expect that in this case the small $a_{bb}$ can hardly change the spectrum obtained with zero $a_{bb}$ (see Fig.\ref{fig3}(b2)).




\section{Summary and discussion}

In summary, in this work we have presented an extensive study of using mass imbalance to enhance the Efimov correlation in Bose polarons, following the main idea illustrated in our earlier work\cite{Sun}. We show that a large mass ratio between bosons and the impurity can facilitate the visualization of Efimov signatures in the spectral response of Bose polarons, and meanwhile, it can reduce the relative spectral width and therefore support a more well-defined quasi-particle behavior in the corresponding system. Moreover, taking the realistic Li-Cs system we study the effect of a finite background boson-boson interaction  to the polaron spectrum. It is shown that the Efimov signatures associated with the second lowest Efimov trimer still persists, and the signal can be even pronounced if lowering the boson density to enable an enhanced hybridization between the Efimov branch and the attractive polaron branch. These results demonstrate the novel Efimov correlation and its visible effect in the many-body environment, which hopefully can be directly probed in current cold atoms experiments of Li-Cs and Li-Rb Bose polarons.

In this work, the spectra in Figs.(\ref{fig3},\ref{fig4}) are obtained in the framework of high temperature virial expansion. 
It is thus remarkable that even at such high temperature the three-body effect can produce so significant effect. These rigorous results in the high temperature regime can serve as a benchmark for studying the polaron physics as reducing the temperature to, for instance, the quantum degeneracy and even the Bose condensation regime. 


Our results also shed light on the Fermi polarons with dominated three-body correlations, which can be associated with the formation of universal trimers\cite{KM} or Efimov trimers\cite{Petrov}. We remark that to enable the visibility of Efimov signatures in the spectral response of Fermi polarons, the general requirement of $l_T\sim d$ will still hold true. However, depending on the property of underlying Fermi system, for instance, it can be a Fermi superfluid\cite{Nishida, Cui1} or with spin-orbit coupling\cite{HuHui, Cui2}, the three-body correlation may manifest itself quite differently in the spectral response, which will be left for future studies. 

{\it Acknowledgements.} We thank the Supercomputer Center in Guangzhou for computational support. This work is supported by the National Natural Science Foundation of China (No. 11622436, 11374177, 11421092, 11534014, 11325418), the National Key Research and Development Program of China (No. 2016YFA0301600), and Tsinghua University Initiative Scientific Research Program.

\end{document}